\documentclass[10pt,twocolumn,superscriptaddress,fleqn,floatfix,showpacs]{revtex4-1}
\usepackage{natbib}
\usepackage{graphicx,color}
\usepackage{dcolumn}
\usepackage{bm}
\usepackage{placeins}
\usepackage{multirow}
\usepackage{amssymb}
\usepackage{amsfonts}
\usepackage{amsmath}
\usepackage{makeidx}
\usepackage[bookmarksnumbered,colorlinks,bookmarks,breaklinks=true,linkbordercolor={1 1 1},linktocpage,citecolor=blue]{hyperref}
\usepackage{subfigure}
\usepackage{fancyhdr}
\usepackage{array}
\usepackage{longtable}
\usepackage{verbatim}
\setlongtables
\usepackage[activate={true,nocompatibility},final,tracking=true,kerning=true,spacing=true,factor=1100,stretch=10,shrink=10]{microtype}
\usepackage[normalem]{ulem}
\newcommand{\added}[1]{#1}
\newcommand{\deleted}[1]{#1}

\microtypecontext{spacing=nonfrench}
\newcommand\T{\rule{0pt}{2.6ex}}
\newcommand\B{\rule[-1.2ex]{0pt}{0pt}}

\def\etals{et al.}
\def\ql{``}
\def\qr{''\hspace{0.5mm}}
\def\qrs{''}

\def\TH{$^{232}$Th}

\def\UT{$^{235}$U}
\def\UF{$^{238}$U}

\def\PU{$^{239}$Pu}

\newcommand{\nuc}[2]{\ensuremath{{}^{#2}\text{#1}}}

\begin{document}

\title{
Predicting the optical observables for nucleon scattering on even-even actinides
}

\author{D.\thinspace S.\thinspace Martyanov}
\author{E.\thinspace Sh.\thinspace Soukhovitski\~{\i}}
\affiliation{Joint Institute for Energy and Nuclear Research, 220109, Minsk-Sosny, Belarus}
\author{R.\thinspace Capote}
\email[Corresponding author, electronic address:~]{\underline{r.capotenoy@iaea.org}}
\affiliation{NAPC--Nuclear Data Section, International Atomic Energy Agency, Vienna
A-1400, Austria}
\author{J.\thinspace M.\thinspace Quesada}
\affiliation{Departamento de F\'{\i}sica At\'{o}mica, Molecular y Nuclear, Universidad de
Sevilla, Ap.1065, E-41080 Sevilla, Spain}
\author{S.\thinspace Chiba}
\affiliation{Research Laboratory for Nuclear Reactors, Tokyo Institute of Technology
2-12-1-N1-9 Ookayama, Meguro-ku, Tokyo 152--8550, Japan}

\begin{abstract}
Previously derived Lane consistent dispersive coupled-channel optical model for nucleon scattering on \nuc{Th}{232} and \nuc{U}{238} nuclei is extended to describe scattering on even-even actinides with $Z=$90--98. A soft-rotator-model (SRM) description of the low-lying nuclear structure is used, where SRM Hamiltonian parameters are adjusted to the observed collective levels of the target nucleus. SRM nuclear wave functions (mixed in $K$ quantum number) have been used to calculate coupling matrix elements of the generalized optical model. The \ql effective\qr deformations that define inter-band couplings are derived from SRM Hamiltonian parameters. Conservation of nuclear volume is enforced by introducing a dynamic monopolar term to the deformed potential leading to additional couplings between rotational bands. Fitted static deformation parameters are in very good agreement with those derived by Wang and collaborators using the Weizs{\"a}cker-Skyrme global mass model (WS4), allowing to use the latter to predict cross section for nuclei without experimental data. A good description of scarce \ql optical\qrs experimental database is achieved. SRM couplings and volume conservation allow a precise calculation of the compound-nucleus formation cross sections, which is significantly different from the one calculated with rigid-rotor potentials coupling the ground-state rotational band. Derived parameters can be used to describe both  neutron and proton induced reactions.
\end{abstract}

\pacs{11.55.Fv, 24.10.Ht}
\date{\today }
\maketitle

\section{Introduction}
Actinide nuclei are a major long-term radiological concern in nuclear reactor waste, and their neutron-induced cross sections are very important for safety calculations of advanced reactor systems. However, experimental information for actinides is rather scarce if we exclude so called major actinides (\UT, \UF, \PU, and \TH). Therefore, actinide cross sections for applications should be estimated using theoretical predictions with models that have been \ql calibrated\qr to existing data of better measured major actinides.

Optical model potential (OMP) based on coupled channels calculations is the key ingredient for such predictions, providing \added{total neutron cross sections, direct elastic and inelastic nucleon scattering cross sections and their angular distributions for $(n,n)$-, $(p,n)$- and $(p,p)$-reactions,} compound  nucleus (CN) formation cross section and nucleon transmission coefficients. The latter are used for \added{nucleon-induced reaction calculations with} pre-equilibrium and equilibrium statistical decay models.

Regional optical models based on rigid rotor coupling of the ground-state rotational band (e.g., see Refs.~\cite{Soukhovitskii:04,casoqu05,Capote:08} and references therein) have been traditionally used to calculate optical observables of actinides \added{for nucleon induced reactions}.
Needs of more accurate data for fast reactors require improving the description of scattering data at incident neutron energies from a few keV up to 5--6 MeV to cover the region with the maximum yield of fission neutrons. While the energies of excited states of the ground-state band of even-even actinides below 500~keV are well described by a rigid rotor model, above 500~keV several vibrational bands are observed that need to be considered~\cite{Chan:1982a,Chan:1982b,Sheldon:1986,Kawano:94}.

A dispersive and Lane consistent OMP with extended couplings was derived for coupled-channels calculations of \added{nucleon-induced reactions on} major actinides - \nuc{Th}{232}, \nuc{U}{233}, \nuc{U}{235}, \nuc{U}{238}, and \nuc{Pu}{239} \cite{PRCextendedCoupling}. Derived potential addressed short-comings of rigid-rotor potentials. Rotational bands were built on vibrational bandheads for even-even targets including both axial and nonaxial dynamical deformations. These additional excitations were introduced as a perturbation to the underlying axially symmetric rigid-rotor structure of the ground-state rotational band. However, the inter-band coupling strengths were not predicted but fitted to available experimental data~\cite{PRCextendedCoupling}. Additionally, the nuclear volume was not conserved for introduced vibrational excitations.

\begin{table*}[!tbp]
\vspace{-3mm}
\caption{Hamiltonian parameters of the soft-rotator model for selected even-even actinides. \added{All parameters have no dimension (except $\hbar\omega_0$ which is given in MeV.}}
\begin{tabular}{l|c|ccccccc}
\hline
\multirow{2}{*}{Target}&Number of fitted&\multicolumn{7}{c}{SRM Hamiltonian parameter values}
\rule{0pt}{2.6ex} \\
\cline{3-9}
\T
 & levels (bands) & $\hbar\omega_0$\added{, MeV} & $\mu_{\beta_2}$ & $\mu_{\gamma}$ & $\gamma_0$ & $B_{32}$ & $\mu_{\beta_3}/\beta_{30}$ & $\delta$ \B \\
      \hline
 \T
 \nuc{Th}{228}                  & 31 (5) & 0.909 & 0.243 & 0.367 & 0.263 & 0.180 & 1.00  & 5.61  \\
 \nuc{Th}{230}                  & 29 (4) & 0.799 & 0.244 & 0.808 & 0.288 & 0.176 & 0.93  & 10.20 \\
 \nuc{Th}{232}                  & 31 (5) & 0.702 & 0.295 & 0.277 & 0.259 & 0.224 & 0.72 & 12.09 \\
 \nuc{U}{232}                   & 25 (5) & 0.927 & 0.235 & 0.589 & 0.282 & 0.249 & 0.67  & 10.80 \\
 \nuc{U}{234}                   & 28 (5) & 1.040 & 0.221 & 0.338 & 0.259 & 0.256 & 0.47  & 15.20 \\
 \nuc{U}{236}                   & 21 (4) & 1.150 & 0.220 & 0.327 & 0.259 & 0.307 & 0.76  & 12.00 \\
 \nuc{U}{238}                   & 27 (5) & 0.979 & 0.224 & 0.292 & 0.234 & 0.217 & 0.96  & 13.60 \\
 \nuc{Pu}{238}                  & 18 (5) & 1.220 & 0.215 & 0.324 & 0.256 & 0.336 & 0.90  & 10.30 \\
 \nuc{Pu}{240}                  & 26 (5) & 1.090 & 0.232 & 0.381 & 0.262 & 0.397 & 0.79  & 9.82  \\
 \nuc{Pu}{242}                  & 13 (3) & 1.110 & 0.247 & 0.423 & 0.257 & 0.491 & 0.81  & 11.40 \\
 \nuc{Pu}{244}                  & 12 (2) & 1.110 & 0.266 & 0.413 & 0.231 & 0.575 & 0.67  & 11.40 \\
 \nuc{Cm}{246}                  & 19 (5) & 1.270 & 0.234 & 0.312 & 0.259 & 0.509 & 0.72  & 18.30 \\
 \nuc{Cm}{248}                  & 17 (4) & 1.080 & 0.252 & 0.340 & 0.269 & 0.509 & 0.93  & 15.70 \\
 \nuc{Cf}{250}                  & 14 (5) & 1.260 & 0.202 & 0.260 & 0.227 & 0.252 & 0.78  & 23.20 \\
\hline
\end{tabular}
\label{table:SRMHamiltonian}
\vspace{-4mm}
\end{table*}
The soft rotator model (SRM) of nuclear structure has been successfully applied in coupled-channels  optical model analyses for \added{nucleon induced reactions on} many nuclei \cite{porsuk96,porsuk98,Lee:2011}. However, only recently we were able to derive a \textit{dispersive} coupled-channels optical model potential for actinides based on soft-rotator couplings \cite{ND2016volumeConservation,ND2016saturation} with volume conservation and saturated coupling scheme. The use of the SRM allowed to derive the inter-band coupling strengths from the low-lying nuclear structure of the even-even target nucleus as shown for \nuc{Th}{232} and \nuc{U}{238} in Ref.~\cite{ND2016volumeConservation}. Saturated couplings \cite{ND2016saturation} were also shown to be important for an accurate prediction of the compound-nucleus formation cross sections, which is a key quantity (though not directly measurable) for statistical decay model calculations. Moreover, total cross section differences of \nuc{Th}{232} and \nuc{U}{238} were well described~\cite{ND2016volumeConservation}, which is a very stringent test for the isovector component of the potential. Therefore, the same OMP can be used to describe nucleon-induced reactions on target nuclei with different number of neutrons and protons.

The aim  of this work is to show that, as a further step, the developed formalism can be extended to \added{describe nucleon induced reactions on} other even-even actinides with incomplete experimental information on their level schemes and/or optical model observables. Previously derived OMP~\cite{ND2016volumeConservation,ND2016saturation} will be combined with comprehensive level schemes estimated using a soft-rotator Hamiltonian, and equilibrium nuclear deformations from global mass models will be used \cite{FRDM2012,WS4}.

\section{Optical model potential with multiple band coupling}
A dispersive coupled-channel optical model with extended couplings based on the SRM~\cite{ND2016volumeConservation,ND2016saturation} has been implemented into the OPTMAN code~\cite{sochiw05,somoch04} \added{to calculate cross section for nucleon induced reactions}. A Lane consistent formulation of the generalized optical model \cite{Quesada:2007} is used with dispersive integrals calculated analytically~\cite{CPC,PRC-disp}. A consistent estimation of the CN formation cross section of even-even targets typically requires the coupling of \added{ground state (}GS\added{)} band levels up to $10^+$ and levels of rotational bands built on octupole, quadrupole $\beta$- and $\gamma$-vibrational excitations, and nonaxial ($K\approx2$) bands~\cite{ND2016saturation}.

The main assumption of this work is that the dispersive OMP used to describe nucleon scattering on even-even actinides is essentially independent of the nuclear structure. The individual nuclear structure is accounted for by the parameters of the corresponding SRM Hamiltonian that properly describes the low-lying collective level scheme (including the multi-band coupling strengths) and, of course, by the individual equilibrium deformation parameters and Fermi energies. The SRM \cite{SRM2} nuclear Hamiltonian parameters had been fitted for even-even actinides with $Z=$90--98 that feature sufficient level data~\cite{Levels228,Levels230,Levels232,Levels234,Levels236,Levels238,Levels240,Levels242,Levels244,Levels246,Levels248,Levels250}. Details of the fitting method and a related discussion of obtained parameters will be published elsewhere. Derived SRM Hamiltonian parameters are listed in Table~\ref{table:SRMHamiltonian}.

Having determined the SRM Hamiltonian parameters, it is possible to use the OMP from Ref.~\cite{ND2016volumeConservation} for predicting the optical observables \added{of nucleon-induced reactions} on even-even actinides of interest, the majority of which have very scarce or absent \ql optical\qr~experimental data. However, we still need to define the GS nuclear static (equilibrium) deformation parameters. For those nuclei, where experimental data are abundant, deformation parameters $\beta_2,\,\beta_4$ and $\beta_6$ are fitted to the available data (which is labeled in figures below as \ql nd2016\added{ (best fit)}\qr). GS equilibrium deformations estimated within global nuclear mass models -- the Finite Range Droplet Model (FRDM2012) \cite{FRDM2012} and the Weizs{\"a}cker-Skyrme (WS4) model \cite{WS4} -- can also be used to predict optical observables if no experimental data are available to fit the deformations. Global equilibrium deformations are labeled in figures below as \ql nd2016+FRDM\qrs~and \ql nd2016+WS4\qrs, respectively.

\begin{table}[!hb]
\vspace{-4mm}
\caption{Static deformations fitted to cross section data for nuclei with relatively abundant experimental data.}
\begin{tabular}{l|ccc}
\hline
\T Target & $\beta_2$ & $\beta_4$ & $\beta_6$ \B \\
\hline                                                             
\T \nuc{Th}{232}  & ~0.201 & ~0.067 & -0.007 \\                    
   \nuc{U}{238}   & ~0.221 & ~0.056 & -0.001 \\                    
   \nuc{Pu}{240}  & ~0.212 & ~0.085 & -0.027 \\                    
   \nuc{Pu}{242}  & ~0.213 & ~0.040 & -0.016 \\                    
\hline
\end{tabular}
\label{table:deformations}
\vspace{-2mm}
\end{table}
Nuclei listed in Table~\ref{table:deformations} are the only even-even actinides with measured optical data above the resonance energy region. Therefore, these four nuclei can be used as a natural benchmark to check the predictive power of FRDM2012 and WS4 deformation parameters against the best-fit deformation parameters listed in Table~\ref{table:deformations}.
\begin{table*}[!thb]
\vspace{-3mm}
\caption{Energies of collective excited levels ordered by rotational bands for selected even-even $Z=$90--98 isotopes. Level energies which are not available in \deleted{ENSDF} evaluated \added{nuclear structure} data~\cite{Levels228,Levels230,Levels232,Levels234,Levels236,Levels238,Levels240,Levels242,Levels244,Levels246,Levels248,Levels250} were estimated by the SRM and are marked by \added{asterisk}\deleted{*}.}
\begin{tabular}{r|llllllllllllll}
\hline
\T {$I^\pi$} &\nuc{Th}{228} &\nuc{Th}{230} &\nuc{Th}{232} &\nuc{U}{232} &\nuc{U}{234} &\nuc{U}{236} &\nuc{U}{238} &\nuc{Pu}{238} &\nuc{Pu}{240} &\nuc{Pu}{242} &\nuc{Pu}{244} &\nuc{Cm}{246} &\nuc{Cm}{248} &\nuc{Cf}{250}  \\
\hline \multicolumn{15}{l}{GS rotational band with positive parity} \rule{0pt}{2.6ex} \\ \hline
\T
$2^+$ & 0.0578  & 0.0532   & 0.0494  & 0.0476  & 0.0435  & 0.0452  & 0.0449 & 0.0442  & 0.0428  & 0.0445  & 0.0442   & 0.0429  & 0.0434  & 0.0427  \\
$4^+$ & 0.1868  & 0.1741   & 0.1621  & 0.1566  & 0.1434  & 0.1495  & 0.1484 & 0.1460  & 0.1417  & 0.1473  & 0.1550   & 0.1420  & 0.1436  & 0.1419  \\
$6^+$ & 0.3782  & 0.3565   & 0.3333  & 0.3227  & 0.2961  & 0.3098  & 0.3072 & 0.3034  & 0.2943  & 0.3064  & 0.3179   & 0.2949  & 0.2981  & 0.2962  \\
$8^+$ & 0.6225  & 0.5938   & 0.5569  & 0.5411  & 0.4970  & 0.5223  & 0.5181 & 0.5136  & 0.4974  & 0.5181  & 0.5350   & 0.5005  & 0.5050  & 0.5000  \\
$10^+$& 0.9118  & 0.8793   & 0.8268  & 0.8059  & 0.7412  & 0.7823  & 0.7759 & 0.7735  & 0.7474  & 0.7786  & 0.8024   & 0.7533  & 0.7607  & 0.7486* \\
\hline \multicolumn{15}{l}{Octupolar rotational band with negative parity} \rule{0pt}{2.6ex} \\ \hline
\T
$1^-$ & 0.3280  & 0.5082   & 0.7144  & 0.5632  & 0.7863  & 0.6876  & 0.6801 & 0.6051  & 0.5973  & 0.7805   & 0.8961* & 1.2498  & 1.0490  & 1.1755  \\
$3^-$ & 0.3961  & 0.5718   & 0.7744  & 0.6290  & 0.8493  & 0.7442  & 0.7319 & 0.6614  & 0.6489  & 0.8323   & 0.9570  & 1.3004  & 1.0940  & 1.2369  \\
$5^-$ & 0.5192  & 0.6866   & 0.8838  & 0.7468  & 0.9626  & 0.8481  & 0.8266 & 0.7632  & 0.7423  & 0.9270   & 1.0680  & 1.3970  & 1.1720  & 1.3361* \\
$7^-$ & 0.6955  & 0.8521   & 1.0429  & 0.9152  & 1.1253  & 0.9996  & 0.9663 & 0.8923* & 0.8781  & 1.0610*  & 1.2063  & 1.5362* & 1.2877* & 1.4783* \\
$9^-$ & 0.9208  & 1.0653   & 1.2496  & 1.1311  & 1.3356  & 1.1984  & 1.1507 & 1.0688* & 1.0568  & 1.2339*  & 1.3953  & 1.7165* & 1.4395* & 1.6621* \\
\hline \multicolumn{15}{l}{$\beta$-rotational band with positive parity} \rule{0pt}{2.6ex} \\ \hline
\T
$0^+$ & 0.9386  & 0.7987*  & 0.7306  & 0.9273* & 1.0445  & 1.1511* & 0.9930 & 1.2287  & 1.0895  & 1.1136*  & 1.1141* & 1.2893  & 1.0840  & 1.2666  \\
$2^+$ & 0.9795  & 0.8583*  & 0.7742  & 0.9676  & 1.0853  & 1.1989* & 1.0373 & 1.2642  & 1.1310  & 1.1591*  & 1.1591* & 1.3176  & 1.1260  & 1.2966  \\
$4^+$ & 1.0748  & 0.9917*  & 0.8730* & 1.0982  & 1.1952* & 1.3090* & 1.1308 & 1.3785* & 1.0895* & 1.2649*  & 1.2647* & 1.3792  & 1.2220  & 1.4089* \\
$6^+$ & 1.3189* & 1.1869*  & 1.0233* & 1.2771* & 1.3575* & 1.4784* & 1.2692 & 1.5445* & 1.3982* & 1.4297*  & 1.4311* & 1.5688* & 1.3873* & 1.5702* \\
\hline \multicolumn{15}{l}{$\gamma$-rotational band with positive parity} \rule{0pt}{2.6ex} \\ \hline
\T
$0^+$ & 0.8318  & 0.6349   & 1.0786  & 0.6914  & 0.8099  & 0.9191  & 0.9272 & 0.9415  & 0.8607  & 0.9560   & 1.1016* & 1.1747  & 1.0538* & 1.1542  \\
$2^+$ & 0.8745  & 0.6775   & 1.1217  & 0.7346  & 0.8517  & 0.9579  & 0.9661 & 0.9831  & 0.9003  & 0.9925   & 1.1434* & 1.2105  & 1.0945* & 1.1894  \\
$4^+$ & 0.9685  & 0.7755   & 1.2221  & 0.8331  & 0.9476  & 1.0509  & 1.0564 & 1.0861* & 0.9924  & 1.0903*  & 1.2382* & 1.3069* & 1.1876* & 1.2919* \\
\hline \multicolumn{15}{l}{Non-axial rotational band with positive parity} \rule{0pt}{2.6ex} \\ \hline
\T
$2^+$ & 0.9690  & 0.7814   & 0.7853  & 0.8668  & 0.9267  & 0.9603  & 1.0603 & 1.0285  & 1.1370  & 1.4993*  & 2.5701* & 1.1243  & 1.0490  & 1.0319  \\
$3^+$ & 1.0225  & 0.8257   & 0.8296  & 0.9115  & 0.9684  & 1.0015  & 1.1057 & 1.0699  & 1.1776  & 1.5375*  & 2.6063* & 1.1655  & 1.0940  & 1.0714  \\
$4^+$ & 1.0911  & 0.8836   & 0.8901  & 0.9707  & 1.0238  & 1.0588  & 1.1630 & 1.1258  & 1.2325  & 1.5890*  & 2.6552* & 1.2200  & 1.1430  & 1.1230  \\
\hline
\end{tabular}
\label{table:LevelsI}
\vspace{-3mm}
\end{table*}

For all calculated targets we used the saturated level scheme, which was previously tested for \nuc{U}{238} and \nuc{Th}{232}~\cite{ND2016saturation}. The (coupled) level energies employed in our coupled-channel calculations are listed in Table~\ref{table:LevelsI}. If the needed level energy was not measured experimentally, then the level energy is calculated using the SRM with parameters from Table~\ref{table:SRMHamiltonian}; such cases are marked by \added{asterisk}\deleted{*} in the Table.

\begin{figure*}[!htbp]
\centering
\subfigure[~\nuc{Pu}{240}]{\includegraphics[width=0.95\columnwidth]{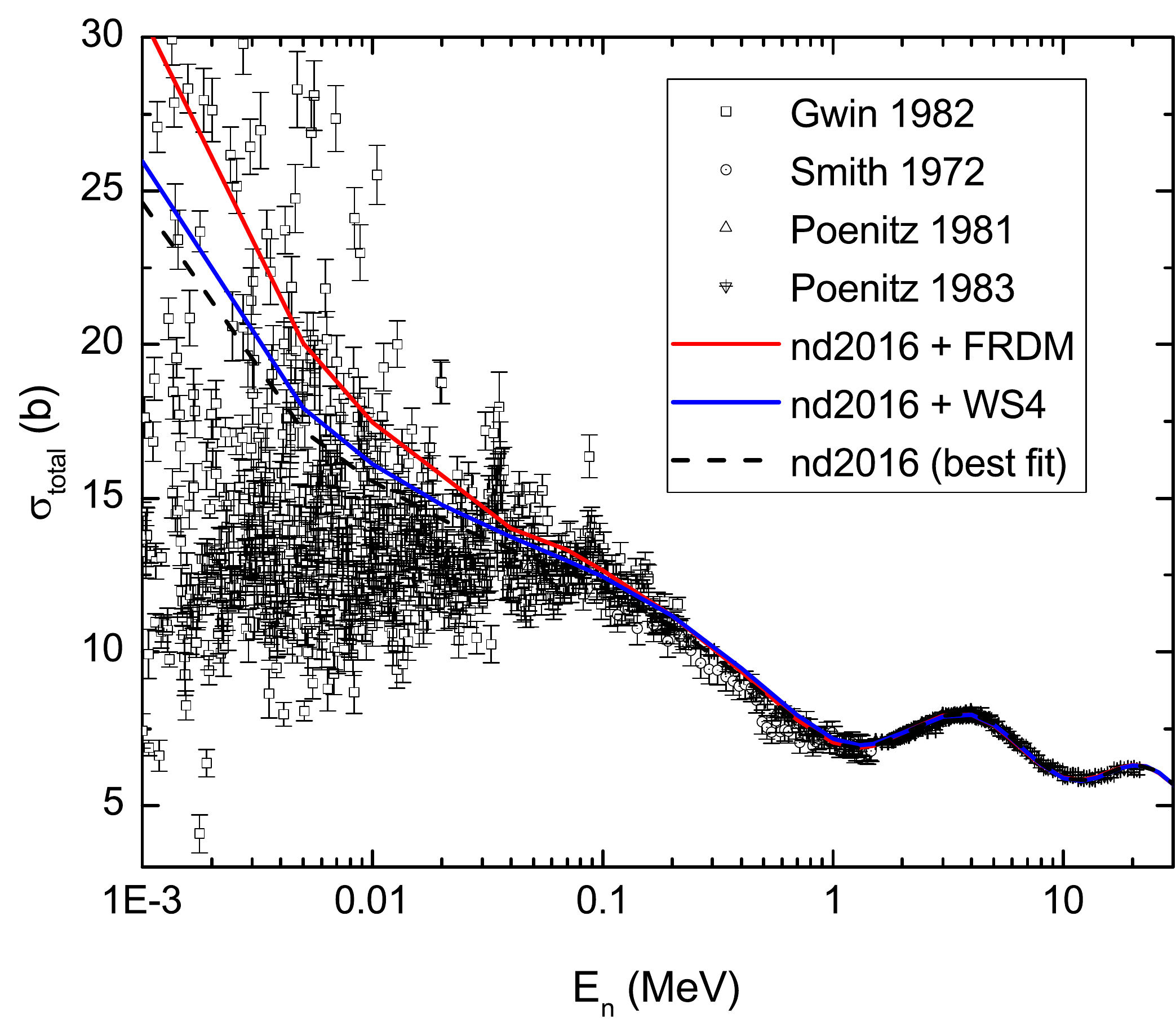}}\hfill
\subfigure[~\nuc{Pu}{242}]{\includegraphics[width=0.95\columnwidth]{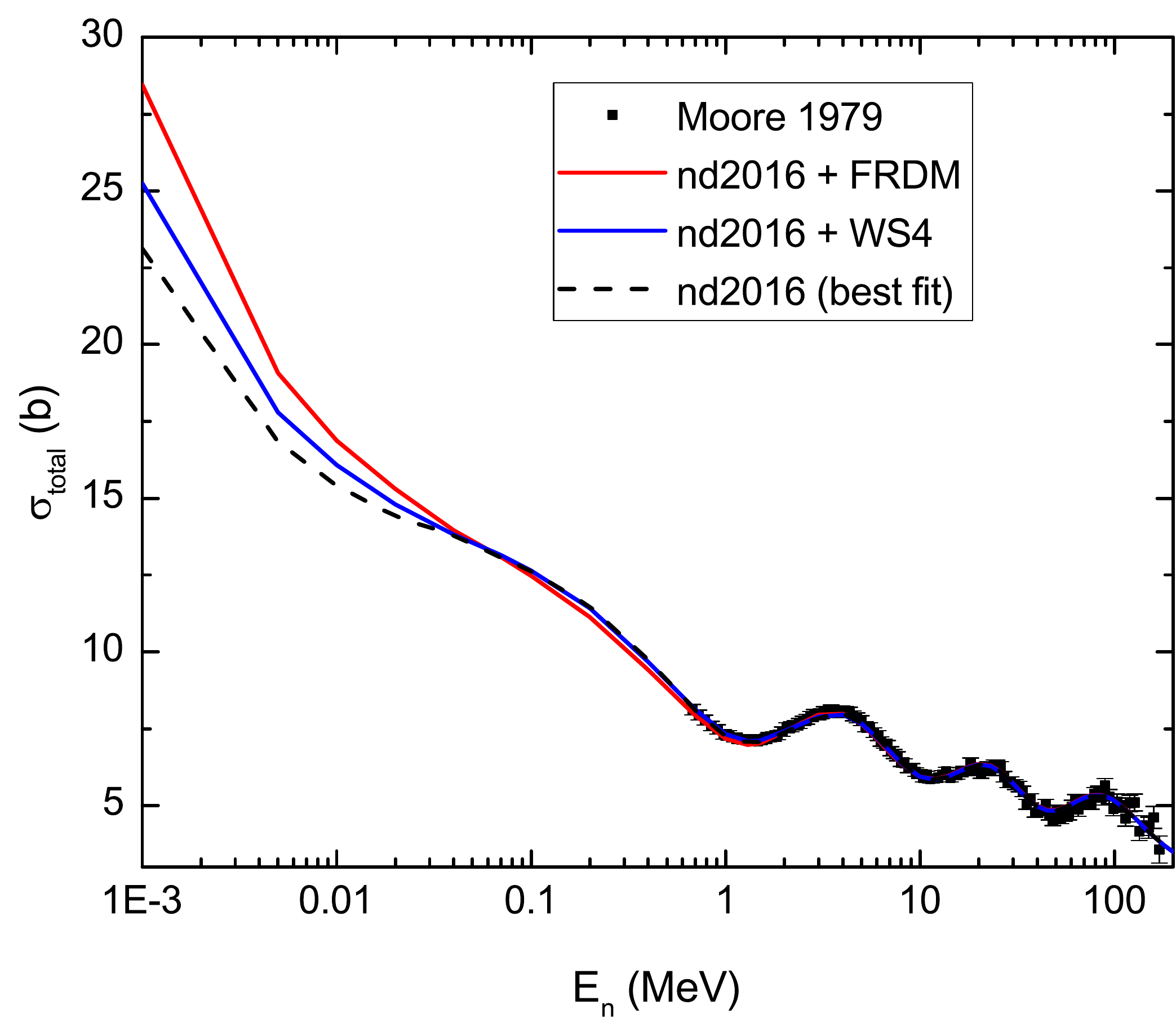}}
\vspace{-3mm}
\caption{(Color online) Total cross sections for \nuc{Pu}{240} and \nuc{Pu}{242}: experimental data vs predicted cross sections using different static deformations.}
\label{fig:TotalXS}
\end{figure*}

\begin{figure}[!thb]
\centering
\includegraphics[width=0.95\columnwidth]{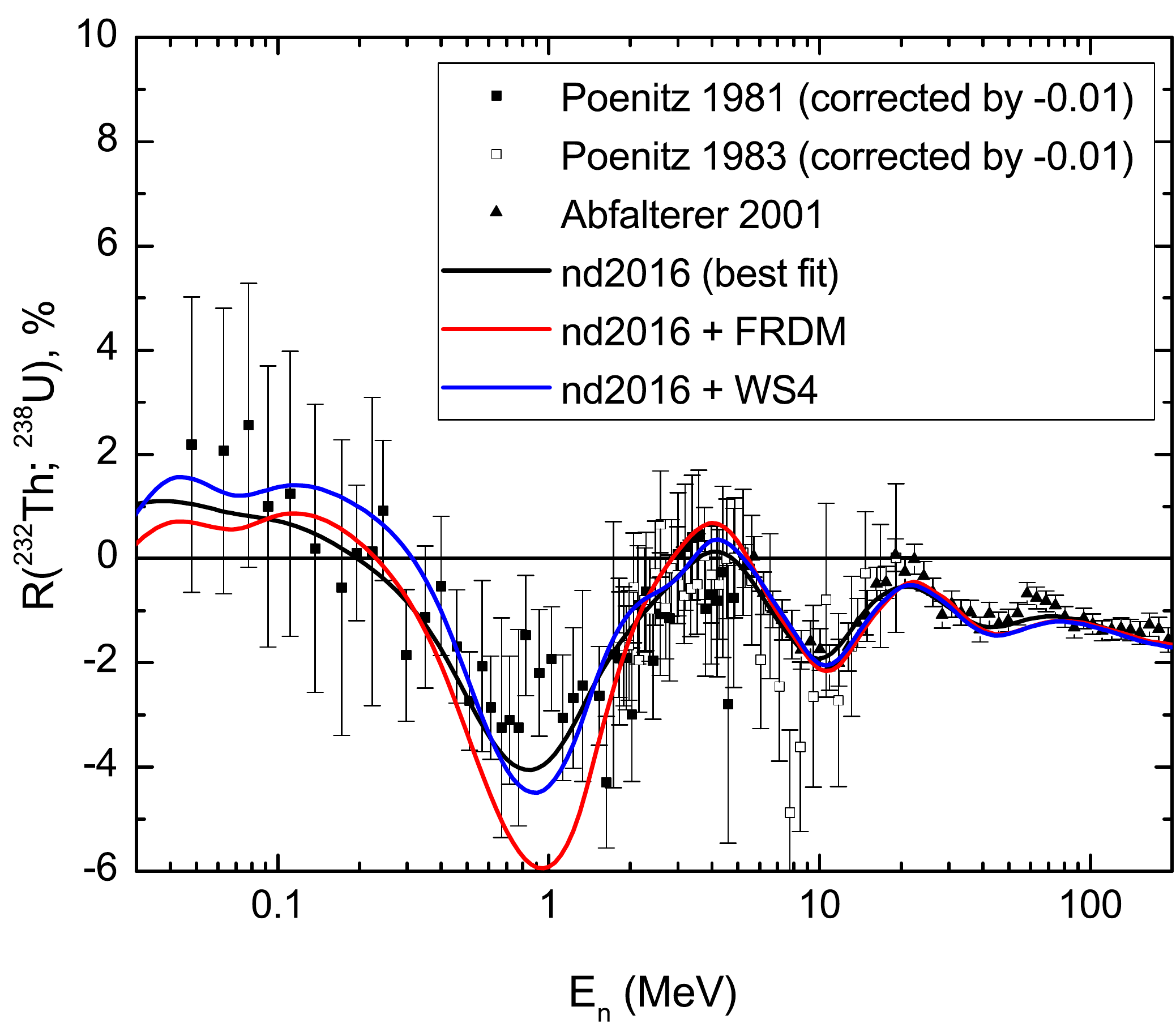}
\caption{(Color online) Calculated total cross section ratio defined as $R(A;B)=2\frac{\sigma_A-\sigma_B}{\sigma_A+\sigma_B}$ for \TH~and \UF~cross sections using different sets of deformations.}
\label{fig:th232-to-u238ratio}
\end{figure}

To validate the proposed approach, we have compared the predicted cross sections with above-mentioned sets of static deformation parameters
$\beta_{2}$, $\beta_{4}$, and $\beta_{6}$ for 
\nuc{Pu}{240} and \nuc{Pu}{242} targets. Fig.~\ref{fig:TotalXS} shows the experimental data vs. the results of using the three different deformation sets (labeled \ql nd2016\added{ (best fit)}\qrs, \ql nd2016+FRDM\qrs, and \ql nd2016+WS4\qrs) in total cross-section calculations.
One can see that using the OMP~\cite{ND2016volumeConservation,ND2016saturation} with SRM coupled levels from Table~\ref{table:LevelsI} combined with either FRDM2012 or WS4 deformation parameters $\beta_{2}$, $\beta_{4}$, $\beta_{6}$ as static deformations leads to rather good reproduction of the experimental data above 0.1~MeV incident neutron energies. Calculated results also agree with results of using the best-fit deformation parameters (\ql nd2016\added{ (best fit)}\qrs). However, there are sizeable differences between different calculations at incident energies below 100~keV, especially if we use FRDM deformations. 
The suggested approach performs very well for the accurately measured energy dependence of the ratio $R($\nuc{Th}{232}$;$\nuc{U}{238}$)$ (defined as $R(A;B)=2\frac{\sigma_A-\sigma_B}{\sigma_A+\sigma_B}$) as shown in Fig.~\ref{fig:th232-to-u238ratio}. Measured ratio is reproduced within experimental uncertainty for best-fit deformation parameters (\ql nd2016~(best fit)\qrs~curve), and also for WS4 deformation parameters (\ql nd2016+WS4\qrs~curve). Predictions with FRDM2012 deformations (\ql nd2016+FRDM\qrs curve) are slightly outside the experimental uncertainty band at incident neutron energy around 1~MeV as seen in Fig.~\ref{fig:th232-to-u238ratio}.

\begin{figure}[!thb]
\centering
\includegraphics[width=0.95\columnwidth]{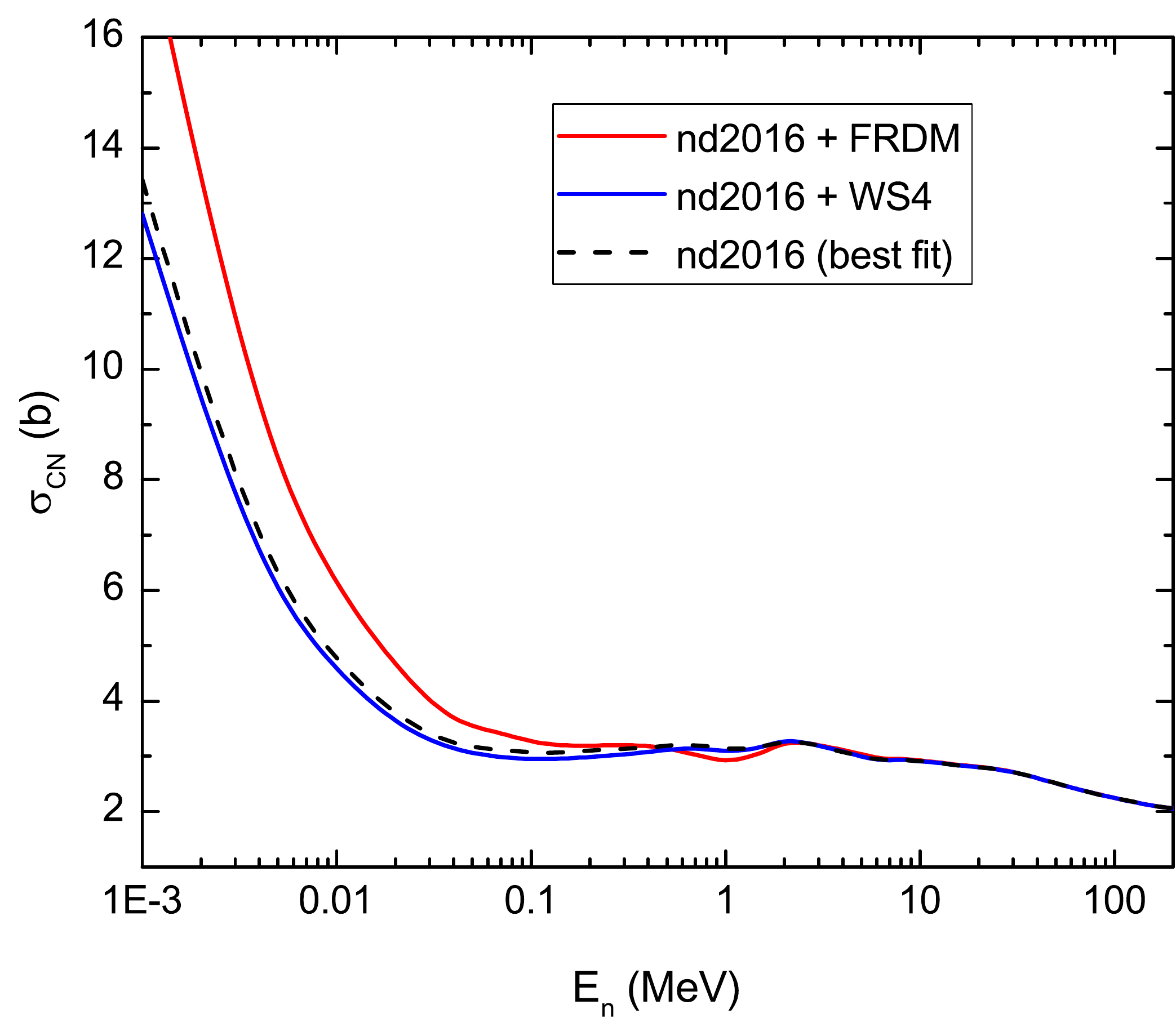}
\caption{(Color online) Predicted compound-nucleus formation cross sections for $n+$\UF~reaction using different static deformations.}
\label{fig:allCN}
\end{figure}


\begin{table}[!tbh]
\vspace{-5mm}
\caption{Calculated $\chi^2$ values for nuclei with abundant experimental data for three different deformation sets: best-fit\added{ (all $\beta_2, \beta_4, \beta_6$ adjusted with values listed in Table~\ref{table:deformations}), and FRDM2012 and WS4 equilibrium deformations predicted by global mass models.}}
\begin{tabular}{l|c|c|c}
\hline
\T
\multirow{2}{*}{Target} & \multicolumn{3}{c}{Calculated $\chi^2$ value}\B \\
\cline{2-4}
\T
    & Best fit ($\beta_2, \beta_4, \beta_6$) &  FRDM2012 \cite{FRDM2012} &  WS4 \cite{WS4} \B\\
\hline                                                                  
\T
\nuc{Th}{232}  & 2.65 & 11.8 & 3.45 \\                    
\T
\nuc{U}{238}   & 1.87 & 14.6 & 2.30 \\                    
\T
\nuc{Pu}{240}  & 0.56 & 1.26 & 0.81 \\                    
\T
\nuc{Pu}{242}  & 0.60 & 1.26 & 0.69 \\                    
\hline
\end{tabular}
\label{table:chi-2}
\vspace{-2mm}
\end{table}



Calculated $\chi^2$ values using available experimental data for neutron and proton induced reactions are given in Table~\ref{table:chi-2} for the three considered deformation sets. If we have experimental data, then we can use it to derive the best-fit values shown in column \ql best~fit~($\beta_2, \beta_4, \beta_6$)\qrs, which guarantee the lowest $\chi^2$.

However, if no experimental data are available, then WS4 deformations result in an overall $\chi^2$ (shown in Column 4 of Table~\ref{table:chi-2}) which is significantly better than the $\chi^2$ obtained for FRDM2012 deformations (shown in Column 3 of Table~\ref{table:chi-2}). Differences in $\chi^2$ are especially large for \TH~and \UF~targets, where many experimental data are available. Therefore, we may conclude that WS4 deformation parameters have a better predictive power to calculate optical model cross sections than those obtained from FRDM model.

Best fit deformations (in nuclides for which they can be fitted due to a sufficient amount of experimental data) reproduce very well the evaluated neutron-strength function values ($S_0$) as shown in Table~\ref{table:SFandDef2}. Calculations with WS4 deformations predict $S_0$-values which are much closer to the data than the ones predicted with FRDM2012 deformations as seen in the same Table. It should be noted that for actinides not having experimental data above the resonance region, but with experimentally determined $S_0$, the accuracy of predicted cross sections could be further improved by adjusting $\beta_2$ to reproduce the measured neutron-strength function $S_0$-value.

\begin{table*}[!tbh]
\caption{Evaluated neutron strength functions from Refs.~\cite{mu06,ripl3,posuma98} are compared with values fitted (best fit) or predicted (FRDM2012 and WS4) using  different sets of deformation parameters for all considered even-even actinides.}
\begin{tabular}{l|ccr@{$\pm$}l|c|c|c}
\hline
\multirow{2}{*}{Target}  & \multicolumn{4}{c|}{Evaluated $S_0$} & \multicolumn{1}{c|}{best fit (ref)}  & \multicolumn{1}{c|}{FRDM2012 \cite{FRDM2012}} &  WS4 \cite{WS4} \T \\
\cline{2-8}
 & BNL \cite{mu06} & RIPL \cite{ripl3} & \multicolumn{2}{c|}{Minsk \cite{posuma98}
 \footnote{The values for this column are taken  from Ref.~\cite{posuma98} if not superseded by values from later reports by Minsk group, in which case the references~\cite{th232rep,u238rep,u232rep,u234rep} are specified}}
 & $S_0$ &  $S_0$ &  $S_0$  \T \\
 \hline
\nuc{Th}{232}& 0.71$\pm$0.04 & 0.84$\pm$0.07 & 0.94&0.07 \cite{th232rep} & 0.93 &  1.14 &  0.80  \T \\
\nuc{U}{238} & 1.29$\pm$0.13 & 1.03$\pm$0.08 & 1.03&0.08 \cite{u238rep}  & 1.01 &  1.39 &  0.96  \\
\nuc{Pu}{240}& 1.11$\pm$0.08 & 1.07$\pm$0.10 & 1.07&0.16                 & 1.03 &  1.43 &  1.12  \\
\nuc{Pu}{242}& 0.92$\pm$0.10 & 0.98$\pm$0.08 & 0.91&0.15                 & 0.89 &  1.30 &  1.05  \\
\hline
 \nuc{Th}{228}&         -     &      -        & \multicolumn{2}{c|}{-}   &  -   &  0.84 &  0.74 \T \\
 \nuc{Th}{230}& 1.75$\pm$0.51 & 1.28$\pm$0.15 & 1.39&0.40                &  -   &  0.96 &  0.70  \\
 \nuc{U}{232} & 0.91$\pm$0.2  & 1.00$\pm$0.20 & 1.17&0.08 \cite{u232rep} &  -   &  1.09 &  0.93  \\
 \nuc{U}{234} & 0.83$\pm$0.11 & 0.85$\pm$0.10 & 0.96&0.13 \cite{u234rep} &  -   &  1.22 &  0.95  \\
 \nuc{U}{236} & 1.03$\pm$0.09 & 0.98$\pm$0.07 & 1.03&0.13                &  -   &  1.31 &  0.97  \\
 \nuc{Pu}{238}& 1.30$\pm$0.30 & 1.08$\pm$0.15 & 1.29&0.27                &  -   &  1.38 &  1.11  \\
 \nuc{Pu}{244}&      -        & 0.90$\pm$0.20 & 1.24&0.52                &  -   &  1.25 &  1.07  \\
 \nuc{Cm}{246}& 0.63$\pm$0.26 & 0.84$\pm$0.15 & 0.91&0.34                &  -   &  1.64 &  1.16  \\
 \nuc{Cm}{248}& 1.22$\pm$0.22 & 1.20$\pm$0.20 & 1.01&0.24                &  -   &  1.38 &  1.04  \\
 \nuc{Cf}{250}&      -        &   -           &\multicolumn{2}{c|}{-}    &  -   &  1.20 &  1.01  \\
\hline
\end{tabular}
\label{table:SFandDef2}
\end{table*}

Fig.~\ref{fig:allCN} shows the comparison of calculated neutron-induced CN formation cross section for neutrons incident on \nuc{U}{238} target. There are no direct CN formation cross-section measurements. We have considered the best fit deformation parameters listed in Table~\ref{table:deformations} to define a reference calculation labelled as \ql nd2016 (best fit)\qrs. Such reference calculation is compared with calculations using the FRDM2012~\cite{FRDM2012} and WS4~\cite{WS4} values of deformation parameters. Results using WS4 deformation parameters agree perfectly with the reference calculation in the whole energy range. Use of FRDM2012 parameters leads to higher CN formation cross sections below 100~keV. Therefore, for prediction of the CN formation cross section above 100~keV we can use both FRDM2012 and WS4 deformations. If CN formation cross sections are needed below 100 keV, it is better to use the global WS4 deformation parameters.

\section{Conclusion}

Previously developed dispersive optical model with extended couplings \cite{PRCextendedCoupling,ND2016volumeConservation}
that consider volume conservation has been extended to predict nucleon-nucleus scattering cross sections on selected even-even actinides.
No additional OMP parameters are needed. SRM Hamiltonian parameters had been obtained from the experimentally  observed low-lying collective
levels or predicted from systematics if levels are not experimentally observed.
Saturated coupling scheme \cite{ND2016saturation} based on SRM description of the nuclear structure
is proposed to calculate the extended couplings. Equilibrium deformations are
fitted to reproduce experimental data when available; otherwise static deformations should be taken from those predicted by global mass models FRDM 2012 \cite{FRDM2012} and WS4 \cite{WS4}. The use of WS4 deformation parameters looks preferable to the use of FRDM2012 deformation parameters, especially to calculate neutron-induced reaction cross sections for incident neutron energies below 100~keV.
A similar extension of the proposed OMP to describe odd actinides is warranted.

\begin{acknowledgments}
This work was partially supported by International Atomic Energy Agency, through the IAEA Research Contract 19263,  by the Spanish Ministry of Economy and Competitivity under Contracts FPA2014-53290-C2-2-P and FPA2016-77689-C2-1-R.
\end{acknowledgments}


\end{document}